\documentclass[aps,pra,superscriptaddress,twocolumn]{revtex4-1}
\usepackage{natbib}
\usepackage{amsmath}
\usepackage{amssymb}
\usepackage{amsthm}
\usepackage{nicefrac}
\usepackage{graphicx}
\usepackage{bm}
\usepackage{url}

\newcommand{\im}[1]{\operatorname{Im}\left[#1\right]}
\newcommand{\re}[1]{\operatorname{Re}\left[#1\right]}

\newcommand{\sym}[1]{\operatorname{Sym}\left[#1\right]}
\newcommand{\asym}[1]{\operatorname{Asym}\left[#1\right]}

\newcommand{\unitv}[1]{\hat{\textbf{#1}}}

\begin{document}
\title{$\mathbb{T}$-Operator Limits on Optical Communication: \\
 Metaoptics, Computation, and Input--Output Transformations}
\author{S. Molesky}
\thanks{Equal contribution}
\author{P. Chao}
\thanks{Equal contribution}
\author{J. Mohajan}
\affiliation{Department of Electrical and Computer Engineering, 
Princeton University, Princeton, New Jersey 08544, USA}
\author{W. Reinhart}
\affiliation{Department of Materials Science and Engineering, 
Pennsylvania State University, University Park, Pennsylvania 16802, USA}
\author{H. Chi} 
\affiliation{Siemens, Princeton, New Jersey 08540, USA} 
\author{A. W. Rodriguez}
\affiliation{Department of Electrical and Computer Engineering, 
Princeton University, Princeton, New Jersey 08544, USA}
\begin{abstract}
 We present an optimization framework based on Lagrange duality and
 the scattering $\mathbb{T}$ operator of electromagnetism to
 construct limits on the possible features that may be imparted to a
 collection of output fields from a collection of input fields, i.e.,
 constraints on achievable optical transformations and
 the characteristics of structured materials as communication channels.
 Implications of these bounds on the performance of representative
 optical devices having multi-wavelength or multi-port
 functionalities are examined in the context of electromagnetic
 shielding, focusing, near-field resolution, and linear computing.
\end{abstract}
\maketitle

As undoubtedly surmised since long before Shannon's pioneering work on 
communication~\cite{verdu1998fifty}, or Kirchhoff's investigation of 
the laws governing thermal radiation~\cite{robitaille2009kirchhoff}, 
physics dictates that there are meaningful limits on how measurable 
quantities may be transferred between senders and receivers 
(collectively \emph{registers}) that apply largely independent of the 
precise details by which transmission is realized. 
The noisy-coding theorem, for instance, proves that probabilistically 
error free message passing is not possible at any rate larger than the 
``channel capacity''~\cite{shannon1948mathematical}; while more 
recently, the possibility of utilizing entanglement as a novel resource 
has motivated the development of a variety of limits on communication in 
general quantum systems, e.g. Refs.~\cite{pendry1983quantum,
caves1994quantum,horodecki2000limits,pirandola2017fundamental}. 
Broadly, even at the coarsest levels of physical description, there is 
generally some notion of invariants, and the existence of such 
quantities implicitly precludes the possibility of realizing complete 
engineering control. 

Equating registers with electromagnetic fields and scattering objects
with channels~\cite{miller2019waves}, using the language of
communication theory, it is thus reasonable to assume that there are
certain channel-based limits on the extent that a collection of fields
may be manipulated via material structuring.  The wave nature of
Maxwell's equations sets a fundamental relation between wavelength and
frequency (for propagation) that cannot be arbitrarily altered by any
realizable combination of material and geometry. Consequently, the
scattered fields generated by any true object cannot be matched to any
freely selected magnitude and phase profile, and so, certain
transformations cannot be achieved with perfect fidelity (e.g. known
limits on light
trapping~\cite{yablonovitch1982statistical,callahan2012solar,
  miroshnichenko2018ultimate}, cloaking
bandwidths~\cite{hashemi2010delay,monticone2016invisibility,
  cassier2017bounds}, delay-bandwidth
products~\cite{caulfield1977optical,tsakmakidis2017breaking,
  mann2019nonreciprocal}, etc.)
\begin{figure}[t!]
 \centering
 \includegraphics[width=1.0\columnwidth]{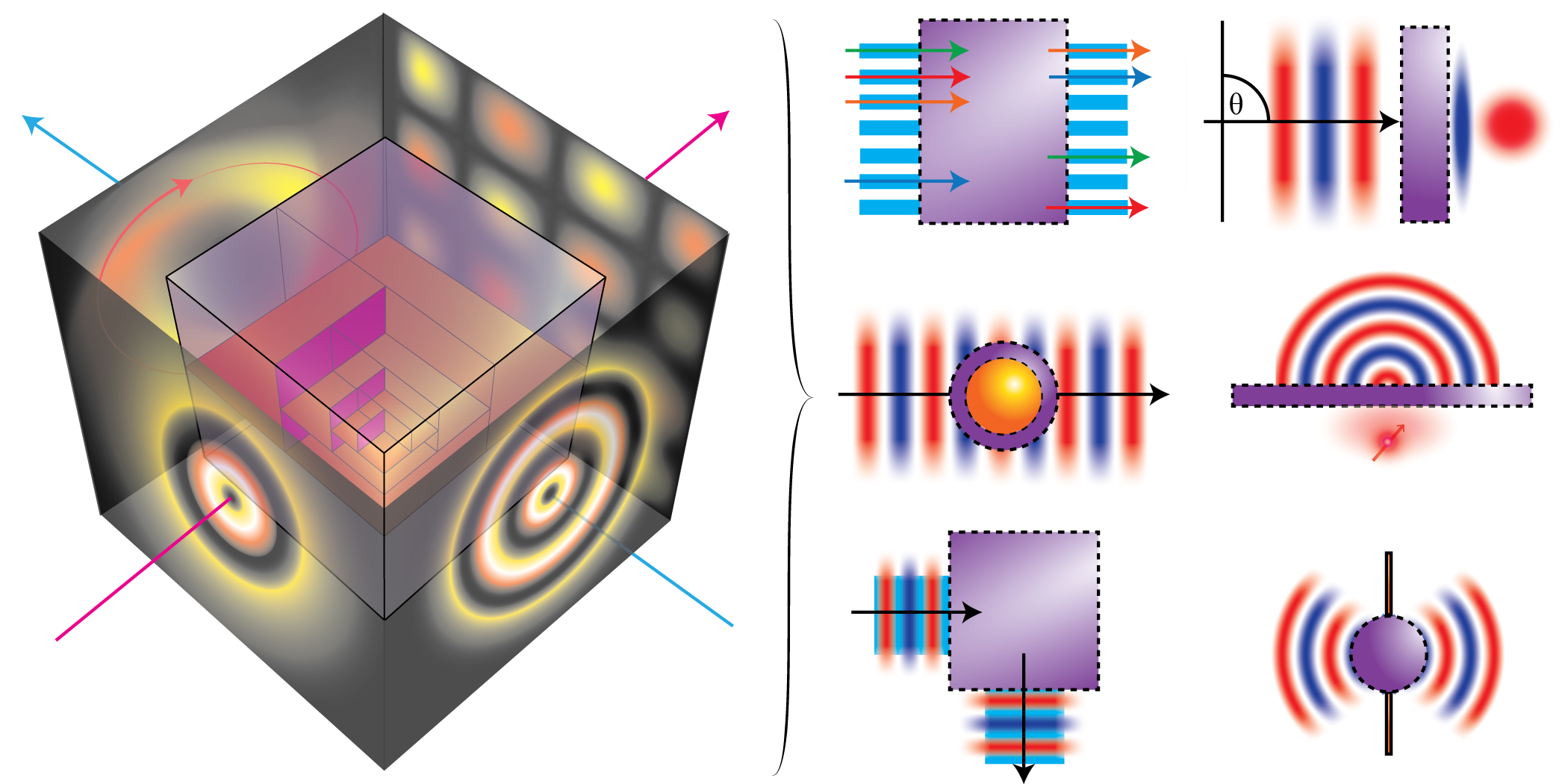}
 \caption{\textbf{Investigation schematic and sample applications.}
  The figure sketches the central question explored in this article:
  given a specified design volume(s) and material(s), how effectively
  can some particular collection of field transformations (described
  as input--output pairs) be realized? The panel on the left
  depicts an abstract optical device, the imitable black-box,
  converting a known set of input electromagnetic field profiles
  into a desired set of output field profiles. The partitions
  interior to the design domain suggest how structural degrees of
  freedom or constraint clusters may be refined to improve design
  performance and estimations of limit performance (bounds),
  respectively. The smaller images to the right depict six (of the
  many) possible applications that can be easily described within
  this framework. Working from left-to-right, top-to-bottom, theses
  illustrations represent a spatial multiplexer, a metaoptic lens, a
  cloaking shell for an enclosed object, a light extractor
  (enhancing radiative emission from a dipolar source), a waveguide
  bend, and a directional antenna. }
\end{figure}

However, in establishing a means of evaluating the potential of
present and future electromagnetic devices to address challenges
requiring complex functionalities, such as artificial neural
networks~\cite{abu1987optical,hughes2019wave,zuo2019all,
  bogaerts2020programmable}, spatial
multiplexing~\cite{richardson2013space,li2017recent, yang2020density}
and computing~\cite{estakhri2019inverse,li2019intelligent,
  rajabalipanah2020space}, the important question is not whether such
fundamental limitations exist, but rather what quantitative
implications can be deduced in a given setting.  Beyond well
established considerations like the need to conserve power in passive
systems and the classical diffraction limit of
vacuum~\cite{giovannetti2009sub}, such specific channel features are
typically unknown~\cite{molesky2020t}.  Moreover, it is seldom clear
what level of possible performance improvement could be sensibly
supposed, even to within multiple orders of
magnitude~\cite{miller2015shape,jin2019material,
  venkataram2020fundamental}.  Architectures possessing impressive
field-transformation capabilities have already been demonstrated, from
free-space (grating)
couplers~\cite{piggott2014inverse,michaels2018inverse} to beam
steering~\cite{yang2018freeform,thureja2020array} and polarization
control~\cite{akgol2018design,chung2020high}, suggesting that
large-scale optimization methods may allow scattering attributes to be
tailored to a far greater degree than what has been seen in past
intuition-based designs. Simultaneously, ramifying from the core ideas
of Lagrange duality and interpreting physical relations as
optimization constraints expounded
below~\cite{lu2010inverse,lu2012objective,miller2016fundamental,
  angeris2019computational}, a string of recent articles on improved
bounds for scattering phenomena (including radiative heat
transfer~\cite{venkataram2020fundamental}, absorbed
power~\cite{molesky2019bounds}, scattered
power~\cite{molesky2020fundamental}, and Purcell
enhancement~\cite{molesky2020hierarchical}) have shown that, in some
cases, only modest improvements over standard designs are even
hypothetically attainable~\cite{gustafsson2019upper,
  venkataram2020casimir,molesky2020fundamental,kuang2020maximal,
  schab2020trade,molesky2020hierarchical,trivedi2020fundamental,
  kuang2020computational,jelinek2020sub,angeris2021heuristic}.

In this article, we demonstrate that this rapidly developing program
for calculating bounds on scattering phenomena can be applied to
determine limits on the accuracy to which any particular set of field
transformations may be implemented, providing an initial exploration
of channel characteristics in the context of classical electromagnetic
(multi-wavelength and multi-port) devices.  The article is broken into
three sections.  In Sec~\ref{priorArt}, a condensed overview of
related work examining channel limits on electromagnetic devices prior
to the articles referenced above is provided.  Section~\ref{techDis}
then summarizes the guiding optimization outlook of the rest of the
article along with current methods for constructing
$\mathbb{T}$-operator (scattering theoretic) bounds.  During this
brief overview, two innovations necessary for handling generic
input--output formulations are introduced: a further generalization on
the variety of operator constraints that can be imposed based on the
definition of the $\mathbb{T}$ operator, beyond the possibility of
local clusters examined in
Refs.~\cite{molesky2020hierarchical,kuang2020computational,
  jelinek2020sub}, and a generalization of the type of vector image
constraints that should be usually considered, as required to enforce
that every transformation in an input--output set references the same
underlying structure.  Finally, Sec.~\ref{applications} provides a
number of exemplary applications of the theory, including studies of
model computational kernels, domain shielding and near-field focusing.
For the chosen loss function (the quadratic distance or two-norm
between the set of desired target and realizable fields), the gap
between calculated bounds and the performance of device geometries
discovered via topology (or ``density'') optimization is regularly
found to be of unit order.  The observed trade-offs between the size
of the design domain, the supposed material response of the device,
the specific transformations (channels) considered, the number of
constraint clusters, and the calculated limits reveal several
intuitive trends. Most notably, achievable performance is strongly
dependent on the nature and number of communication channels, with
larger device sizes and indices of refraction resulting in greater
achievable spatial resolution and transformations, in line with prior
heuristics~[REF]. For the wavelength-scale systems investigated here,
the variability in achievable performance spans multiple orders of
magnitude and, as demonstrated in Sec.~\ref{applications}, tied to the
ability of the formulation to enforce that all channel functionalities
map to a unique (single) optimal geometry.

\section{Channel Limits in Prior Art}
\label{priorArt}
\begin{figure*}[t!]
 \centering
 \includegraphics{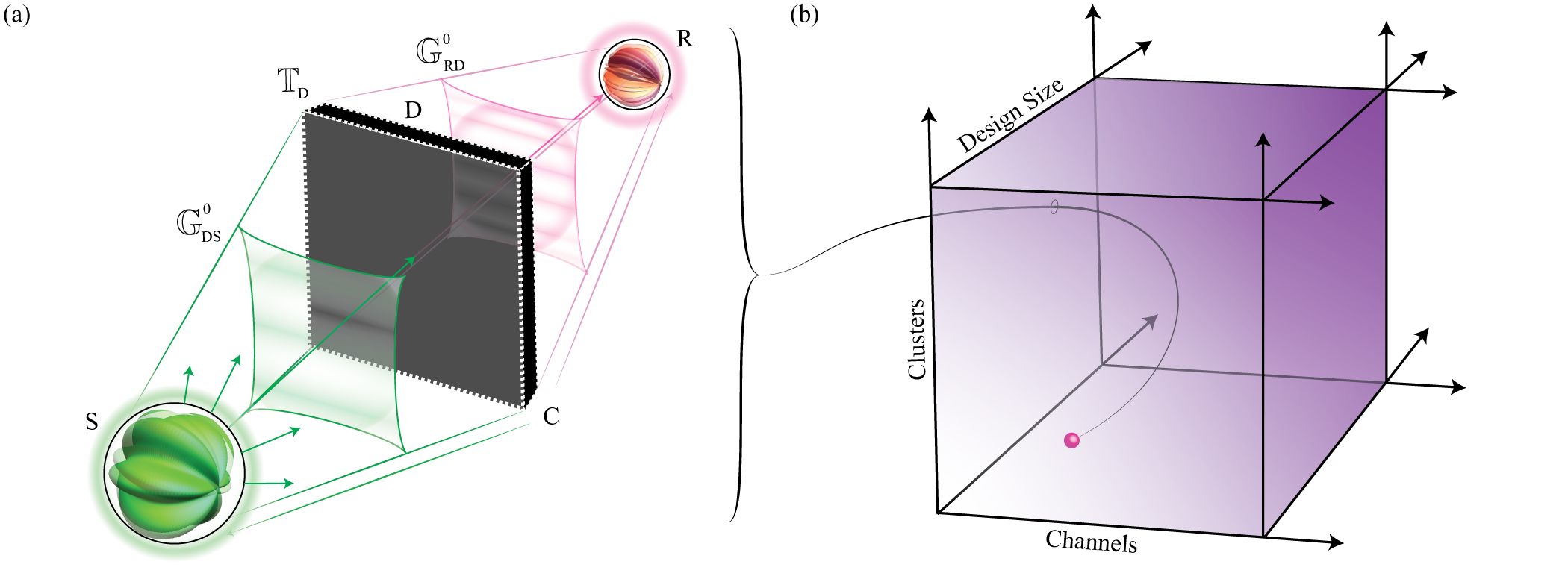}
 \caption{\textbf{Electromagnetic channel limits.} (a) The
  properties of an electromagnetic channel(s) (\textbf{C}) depend on
  the environment connecting a particular (collection of) sender
  (\textbf{S}) and receiver (\textbf{R}) register(s), including
  characteristics like the volume each register occupies, their
  spatial separation, and the possibility of a
  transformation enabling polarizable device (\textbf{D}). Channel
  limits in prior art have been formulated predominately by
  analyzing the individual operators describing each of these
  aspects in isolation (e.g., the rank and pseudo rank of the free
  space Green's function between the sender and receiver volumes
  $\mathbb{G}^{0}_{_{\text{RS}}}$). Here, in contrast, all
  environmental factors are treated simultaneously, accounting for
  the realities of interaction. (b) The difficulty of attaining
  device performance bounds using the formulation presented in
  Sec.~\ref{techDis}, for some predefined material, depends mainly
  on three variables: the number of regional constraints (spatial
  clusters) imposed, the number of channels (simultaneous
  transformations) considered, and the physical volume of the design
  region. The way in which these attributes determine how closely
  the computed bounds approximate achievable device performance is
  generally unknown and problem dependent. For radiative thermal
  emission and integrated absorption, increasing domain size leads
  to increasingly tight bounds when only a single constraint, the
  conservation of real power, is employed~\cite{molesky2019bounds}.
  For radiative Purcell enhancement, recent results suggest that
  predictive bounds for larger design domains require a larger
  number of constraints~\cite{molesky2020hierarchical}. Additional
  regional constraints, or concurrent analysis of additional
  channels, always produces tighter bounds. }
\end{figure*}
Excluding techniques based on Lagrange duality to constrain
electromagnetic design objectives, which are covered in greater detail
in Sec.~\ref{techDis}, three major threads studied in prior art have 
substantially informed the findings and discussion below.

\emph{Decomposition}---Any finite dimensional linear operator (or more
generally any compact operator such as the scattering $\mathbb{T}$
operator of electromagnetism) has a corresponding singular value
decomposition~\cite{rudin2006real}. For almost any example of
practical interest, particularly those with discrete representations,
there are hence corresponding notions of rank and pseudo
rank~\cite{miller2012all,miller2013self,
  miller2017universal,pai2019matrix,miller2019waves}.

\emph{Rank}---Potential field transformations between collections of 
registers are inherently limited by any bounds on maximal rank or pseudo 
rank.
Any set of channels connecting registers that do not overlap in space 
is inherently limited by the rank (pseudo rank) of its associated free 
propagation operators (the Green's 
function)~\cite{miller2007fundamental,miller2013complicated,
miller2015shape,molesky2020fundamental}. 

\emph{Size}---The rank (pseudo rank) characteristics of free propagation 
in electromagnetics depend explicitly on geometry. 
That is, there are cases where possible field transformations are 
strongly limited by the spatial volume occupied by the 
registers~\cite{miller2007fundamental,molesky2019bounds}. 

The main difference between past studies of channel limits utilizing 
these ideas and the $\mathbb{T}$-operator approach given in 
Sec.~\ref{techDis} lies in the incorporation of additional physical 
constraints concerning the generation of polarization currents (within 
a scattering object) in order to effect a desired transformation. 
More concretely, it is often possible to abstractly describe a device 
in terms of (possibly intersecting) design and observation regions. 
Taking a simplistic description of a near-field microscope as an 
example, the magnifying lens may be thought of as a design volume, 
and the final connection between the optical components and the 
electro--optic readout as an observation region.
(As seen in Fig.~2, the design and observation regions are 
closely linked to the sender and receiver registers in optical 
communication.)
Under this regional decomposition, the end goal of various applications 
can be understood as minimizing the power difference between the 
true total field created in the observation region 
($\big|\textbf{E}^{t}_{o}\big>$)~\cite{angeris2019computational}, 
resulting from a known incident field ($\big|\textbf{E}^{i}\big>$), 
and a given target output ($\big|\textbf{E}^{\diamond}_{o}\big>$): 
\begin{equation}
 \text{min}_{\left|\textbf{J}^{g}_{d}\right>}~
 \lVert \big|\textbf{E}^{t}_{o}\big> - 
 \big|\textbf{E}^{\diamond}_{o}\big>\rVert_{2}^{2}.
 \label{basicObj}
\end{equation}
(Here, subscripts mark domains of definitions and superscripts label 
different types of fields.) 
Supposing that polarizable media exists only within the design 
region (the device), following the notation of the upcoming section, 
in terms of the generated polarization field 
$\big|\textbf{J}^{g}_{d}\big>$ Eq.~\eqref{basicObj} becomes 
\begin{equation}
 \text{min}_{\left|\textbf{J}^{g}_{d}\right>}~
 \lVert \big|\textbf{E}^{i}_{o}\big> - 
 \big|\textbf{E}^{\diamond}_{o}\big> + i\frac{Z}{k_{o}}
 \mathbb{G}_{od}^{0}\big|\textbf{J}^{g}_{d}\big>
 \rVert_{2}^{2}. 
 \label{simpCurrMin}
\end{equation}
The minimum of Eq.~\eqref{simpCurrMin} with respect to 
$\big|\textbf{J}^{g}_{d}\big>$, 
\begin{equation}
 i\frac{k_{o}}{Z}\left(\mathbb{G}_{od}^{0}\right)^{\dagger}
 \left(\big|\textbf{E}^{\diamond}_{o}\big> - \big|\textbf{E}^{i}_{o}
 \big>\right) 
 = \left(\mathbb{G}_{od}^{0}\right)^{\dagger}\mathbb{G}_{od} 
 \big|\textbf{J}^{g}_{d}\big>, 
 \label{simpCurrSol}
\end{equation}
illustrates the necessity of the three notions stated above. 
First, even if Eq.~\eqref{simpCurrSol} can be satisfied exactly, the 
minimum of Eq.~\eqref{basicObj} may be nonzero if the span of 
the target basis of $\mathbb{G}^{0}_{od}$ does not contain 
$\big|\textbf{E}^{\diamond}_{o}\big> - \big|\textbf{E}^{i}_{o}\big>$. 
Second, since in extending 
Eqs.~\eqref{basicObj}--\eqref{simpCurrSol} from single fields 
to collections nothing in the form of the minimum solution is altered, 
the rank (pseudo rank) of the Hermitian operator $\left(
\mathbb{G}^{0}_{od}\right)^{\dagger}\mathbb{G}^{0}_{od}$ 
sets a limit on the number of channels that can connect (interact) 
with any register contained in the observation domain. 
Third, as highlighted by the explicit domain subscripts, the 
Green's function ($\mathbb{G}^{0}_{od}$) connecting the observation and 
design domains, on which the two previous points are based, 
depends on the geometric features of both regions. 

While of obvious importance, it is clear that these observations do not 
fully encompass the difficulties associated with the larger question of 
realizing a given set of field transformations. 
Directly, without additional constraints, the challenge associated with 
attaining a particular polarization current 
$\left|\textbf{J}^{g}_{d}\right>$ within the design region cannot be 
inferred from Eq.~\eqref{simpCurrSol}. 
To include this information into the minimization of 
Eq.~\eqref{basicObj}, additional physical constraints must be taken 
into account. 
Achieving this aim, without making the resulting problem statement 
computationally infeasible, is precisely the goal of the
$\mathbb{T}$-operator framework presented below. 
\section{Technical Discussion}
\label{techDis}
\subsection{Scattering theory, Lagrange duality, and limits}
Following the sketch provided in Ref.~\cite{molesky2020t} 
(supporting details can be found in Refs.~\cite{tsang2004scattering,
kirsch2009operator,costabel2012essential,kruger2012trace,
colton2019inverse}), there are two superficial distinctions that 
distinguish a scattering theory view of electromagnetics from Maxwell's 
equations. 
First, supposing a local, linear scattering potential, the 
fundamental wave relation from which all field properties are 
derived is
\begin{equation}
 \mathbb{I}_{s} = \mathbb{I}_{s}\left(\mathbb{V}^{-1} - 
 \mathbb{G}^{\text{0}}\right)\mathbb{T}_{s},
 \label{Tformal}
\end{equation}
where the $s$ subscript denotes projection into the scattering object, 
$\mathbb{G}^{\text{0}}$ is the background Green's function of the design 
domain $\Omega_{s}$, throughout taken to be the vacuum Green's function 
(truncated to a problem-specific volume) scaled by 
$k_{o}^{2} = \left(2\pi/\lambda\right)^{2}$ (so that all spatial 
dimensions may be simply defined relative to the wavelength), 
$\mathbb{V}$ is the bulk (spatially local) scattering potential of the 
material forming the design, $\mathbb{I}$ is the identity operator, 
and $\mathbb{T}_{s}$ is the scattering operator, formally defined by 
Eq.\eqref{Tformal}~\footnote{All operators and fields are supposed to be 
frequency dependent, and every stated relation holds at any given 
frequency.}
\footnote{The potential of the scatterer is constructed via the 
explicit spatial projection operator $\mathbb{I}_{s}$, which is defined 
to be identity at spatial locations within the scatterer and zero at all 
other points in the domain $\Omega$, and the implicit spatial projection 
into the scattering object encoded in the definition the scattering 
operator, $\mathbb{T}_{s}$.}.
(Although more general descriptions are possible, it is implicitly 
assumed throughout this article that $\mathbb{V} = \mathbb{I} \chi$, 
where $\chi$ is electric susceptibility of some isotropic medium.)
Second, one of either the electromagnetic field or polarization 
field is singled out as an initial~\footnote{The specification of an 
initial field is essentially equivalent to the introduction of a source, 
or flux boundary condition, in Maxwell's equations.}, and all other 
fields are generated via the scattering operator $\mathbb{T}_{s}$. 
This leads to a division of the theory into two classes: initial flux 
problems, where a far-field incident flux $\big|\textbf{E}^{i}\big>$ is 
assumed, generating a polarization current 
$\left|\textbf{J}^{g}\right> = -\left(i k_{o}/Z\right)\mathbb{T}_{s}
\big|\textbf{E}^{i}\big>$ and scattered field 
$\big|\textbf{E}^{s}\big> =\left(iZ/k_{o}\right) 
\mathbb{G}^{\text{0}}\big|\textbf{J}^{g}\big>$, and initial 
source problems, where a polarization source $\big|\textbf{J}^{i}
\big>$ is assumed, generating an electromagnetic field 
$\left|\textbf{E}^{g}\right> = \left(iZ/k_{o}\right)
\mathbb{G}^{\text{0}}\mathbb{T}_{s}\mathbb{V}^{-1}
\big|\textbf{J}^{i}\big>$ and total current $\big|\textbf{J}^{t}
\big> = \mathbb{T}_{s}\mathbb{V}^{-1}\big|\textbf{J}^{i}\big>$. 

Given that most quantities related to power transfer result from 
inner products between electromagnetic and polarization 
fields~\cite{tsang2004scattering}, many common design objectives 
within scattering theory, encompassing applications varying from 
enhancing the amount of radiation that can be extracted from a quantum 
emitter~\cite{liu2017enhancing,koenderink2017single,cox2018quantum} to 
increasing absorption into a photovoltaic 
cell~\cite{mokkapati2012nanophotonic,sheng2012light,ganapati2013light}, 
are given by sesquilinear forms on the scattering operator 
$\mathbb{T}_{s}$. 
As the central example investigated in this article, continuing the 
discussion of Sec.~\ref{priorArt} and covering all the specific 
applications treated in Sec.~\ref{applications}, suppose that a 
collection of fluxes $\left\{\big|\textbf{E}^{i}_{k}\big>\right\}$, 
with $k$ ranging over some finite indexing set, is incident on a 
scattering design region $\Omega$. 
Assume that the goal of the (input--output) device is to 
create a set $\Big\{\big|\textbf{E}^{\diamond}_{k}\big>
\Big\}$ of target fields within an observation region $\Omega_{o}$, 
marked by a domain subscript $o$ and spatial projection operator
$\mathbb{I}_{o}$, i.e. a collection of register mappings 
$\big|\textbf{E}^{i}_{k}\big> \mapsto 
\big|\textbf{E}^{\diamond}_{k}\big>$. 
Taking the Euclidean norm as a metric of design performance,
\begin{equation}
 \text{dist}\left(\Big\{\big|\textbf{E}^{t}_{k}\big>\Big\},
 \Big\{\big|\textbf{E}^{\diamond}_{k}\big>\Big\}\right) 
 = \sqrt{\sum_{k}\lVert\mathbb{I}_{o}
 \left(\big|\textbf{E}^{t}_{k}\big> 
 -\big|\textbf{E}^{\diamond}_{k}\big>\right)\rVert^{2}_{2}},
 \label{eucEq}
\end{equation}
the equivalent objective (in the sense \eqref{eucEq} and 
\eqref{objEq} have the same minimum)
\begin{equation}
  \text{Obj}\left(\Big\{\big|\textbf{E}^{t}_{k}\big>\Big\},
 \Big\{\big|\textbf{E}^{\diamond}_{k}\big>\Big\}\right) 
 = \sum_{k}\lVert\mathbb{I}_{o}\left(
 \big|\textbf{E}^{t}_{k}\big> 
 -\big|\textbf{E}^{\diamond}_{k}\big>\right)\rVert^{2}_{2}
 \label{objEq}
\end{equation} 
is effectively a sesquilinear form on $\mathbb{T}_{s}$.
\begin{align}
 \text{Obj}=&
 \sum_{k}\big<\textbf{E}^{\diamond}_{k}\big|
 \mathbb{I}_{o}\big|\textbf{E}^{\diamond}_{k}\big> - 
 2\re{\big<\textbf{E}^{\diamond}_{k}\big|\mathbb{I}_{o}
 \big|\textbf{E}^{t}_{k}\big>} + 
 \big<\textbf{E}^{t}_{k}\big|\mathbb{I}_{o}\left|
 \textbf{E}^{t}_{k}\right> 
 \nonumber \\
 =\sum_{k}& \big<\textbf{E}^{\diamond}_{k}\big|
 \mathbb{I}_{o}\big|\textbf{E}^{\diamond}_{k}\big> - 
 2\re{\big<\textbf{E}^{\diamond}_{k}\big|\mathbb{I}_{o}\big|
 \textbf{E}^{i}_{k}\big>} + 
 \big<\textbf{E}^{i}_{k}\big|\mathbb{I}_{o}\big|
 \textbf{E}^{i}_{k}\big> ~
 \nonumber \\
 &+\big<\textbf{E}^{i}_{k}\big|
 \mathbb{T}^{\dagger}_{s}\mathbb{G}^{0\dagger}
 \mathbb{I}_{o}\mathbb{G}^{0}\mathbb{T}_{s} +
 2\sym{\mathbb{I}_{o}\mathbb{G}^{0}\mathbb{T}_{s}}
 \big|\textbf{E}^{i}_{k}\big> 
 \nonumber \\
 &-2\re{\big<\textbf{E}^{\diamond}_{k}\big|
 \mathbb{I}_{o}\mathbb{G}^{0}\mathbb{T}_{s}\big|
 \textbf{E}^{i}_{k}\big>}, 
 \label{fluxAgree}
\end{align}
and the objective of the design amounts to a minimization of
\begin{align}
 \text{Obj}\left(\mathbb{T}_{s}\right) =
 \sum_{k}& \big<\textbf{E}^{i}_{k}\big|\mathbb{T}^{\dagger}_{s}
 \mathbb{G}^{0\dagger}
 \mathbb{I}_{o}\mathbb{G}^{0}\mathbb{T}_{s} +
 2\sym{\mathbb{I}_{o}\mathbb{G}^{0}\mathbb{T}_{s}}
 \big|\textbf{E}^{i}_{k}\big> 
 \nonumber \\
 &-2\re{\big<\textbf{E}^{\diamond}_{k}\big|
 \mathbb{I}_{o}\mathbb{G}^{0}\mathbb{T}_{s}\big|
 \textbf{E}^{i}_{k}\big>}. 
 \label{fluxInOut}
\end{align}
Here, $\sym{\mathbb{O}}$ and $\asym{\mathbb{O}}$ denote symmetric 
(Hermitian) and anti-symmetric (skew-hermitian) parts of $\mathbb{O}$. 
(Other objective forms, e.g. originating from other norms, could also be
considered, but \eqref{fluxInOut} offers many useful simplifications
for both analysis and computation.) 
Acting with $\mathbb{T}_{s}^{\dagger}$ from the left on \eqref{Tformal}, 
giving
$\mathbb{T}^{\dagger}_{s} =
\mathbb{T}^{\dagger}_{s}\mathbb{U}^{\dagger}\mathbb{T}_{s}$ with
$\mathbb{U}^{\dagger} = \mathbb{V}^{-1} - \mathbb{G}^{0}$ so that
$\asym{\mathbb{U}}$ is positive definite, shows that the physical
constraints of scattering theory can also be written as similar
sesquilinear forms. 
As such, any electromagnetic design objective of this input--output 
form (Ref.~\cite{angeris2019computational}) can be cast as a 
quadratically constrained quadratic program~\cite{phan1982quadratically,
bose2015quadratically} (QCQP) by treating $\mathbb{T}_{s}$ as a vector, 
i.e. 
\begin{align}
 &\text{min}_{~\mathbb{T}_{s}}~\text{Obj}
 \left(\mathbb{T}_{s}\right)\nonumber \\
 &\text{such that}~\left(\forall~l,m\right)~
 \big<\textbf{B}_{l}\big|\mathbb{T}_{s}
 \big|\textbf{B}_{m}\big> = 
 \big<\textbf{B}_{l}\big|\mathbb{T}_{s}^{\dagger}\mathbb{U}
 \mathbb{T}_{s}\big|\textbf{B}_{m}\big>,
 \label{optProb}
\end{align}
over some basis~\footnote{Technically, $\mathcal{B}$ 
should be complete, and hence infinite dimensional. 
In practice, however, $\mathcal{B}$ must truncated to some finite 
set in order to apply numerical methods.
A similar approximation is needed in any numeric solver for 
electromagnetics.} for the space of electromagnetic fields over 
the design region $\Omega_{s}$. 
Although QCQP problems are part of the non-deterministic 
polynomial-time (NP) hard complexity class, a number of relaxation 
techniques, as well as a range of well developed solvers, are known to 
often yield accurate solution approximations~\cite{park2017general} 
(that may in fact be exact). 
Principally, letting 
$\mathcal{L}\left(\mathbb{T}_{s},\bm{\lambda}\right) = 
\text{Obj}\left(\mathbb{T}_{s}\right) + \bm{\lambda} 
~\bm{\mathcal{C}}\left(\mathbb{T}_{s}\right)$
be the Lagrangian of Eq.~\eqref{optProb}, with $\bm{\mathcal{C}}
\left(\mathbb{T}\right)$ denoting the collection of imposed 
scattering constraints and $\bm{\lambda}$ the associated set of 
Lagrange multipliers, it always possible to relax Eq.~\eqref{optProb}, 
or any differentiable optimization problem for that 
matter~\cite{boyd2004convex}, to a convex problem by considering the 
unconstrained dual optimization problem
\begin{equation}
 \text{max}_{\bm{\lambda}}~\mathcal{G}\left(\bm{\lambda}\right), 
 \label{dualOpt}
\end{equation}
with $\mathcal{G}\left(\bm{\lambda}\right) = 
\text{min}_{\mathbb{T}_{s}}~\mathcal{L}\left(\mathbb{T}_{s},
\bm{\lambda}\right)$.
Because any true solution of Eq.~\eqref{optProb} requires every 
constraint relation to be satisfied, for any collection of 
multipliers $\bm{\lambda}$, $\mathcal{G}\left(\bm{\lambda}\right)$ 
is always smaller (resp. larger if the optimization objective is to 
maximize some quantity) than any solution of the primal (original) 
optimization. 
Thus, in addition to providing a means of obtaining approximate 
solutions to Eq.~\eqref{optProb}~\cite{park2017general}, solving 
Eq.~\eqref{dualOpt} also determines a bound, or limit, on the 
performance that could be possibly achieved by any realizable device 
geometry. 
By varying the constraints included in \eqref{optProb} these bounds 
can be tailored to encompass select attributes of the channel(s) as 
one wishes, such as qualifications on material composition (via 
$\mathbb{V}$), the available design domain (via truncation of 
$\mathbb{G}^{\text{0}}$), and the actual input fields (via 
$\Big\{\big|\textbf{E}^{i}_{k}\big>\Big\}$). 
Further, although we have not seriously examined the prospect, it is 
also possible to apply similar reasoning to objective forms differing 
substantially from Eq.~\eqref{objEq}. 

\subsection{Mean-field hierarchy (spatially localized constraints)}
Building on the above perspective, as described in greater detail in 
Refs.~\cite{molesky2020hierarchical,angeris2021heuristic}, the 
$\mathbb{T}$ operator provides a natural means of obtaining arbitrarily 
accurate, context specific, mean-field solution approximations. 
The operator relation given by Eq.~\eqref{Tformal} is equally valid 
under evaluation with any linear functional or vector, or composition 
with any other maps, and can hence be ``coarse-grained'' (averaged). 
Expressly, to construct an appropriate mean-field approximation, 
take $\mathbb{P}_{_{\Omega_{c}}}$ to denote a generalized spatial 
projection operator into the spatial cluster (subregion) $\Omega_{c}$, 
with the added freedom that, at any point within $\Omega_{c}$, 
$\mathbb{P}_{_{\Omega_{_{c}}}}$ may transform the field in any way. 
That is, taking the three dimensional nature of the electromagnetic 
and polarization vector fields into account, $\mathbb{P}_{\Omega_{c}}$ 
may be any operator of the form 
\begin{align}
 \mathbb{P}_{_{\Omega_{c}}}\left(\textbf{x},\textbf{y}\right) =
 \begin{cases}
  \overline{\textbf{0}} & \textbf{x} \neq \textbf{y}~\text{or}~
  \textbf{x}\not\in\Omega_{c} \\
  \overline{\textbf{M}}_{\textbf{x}} & \textbf{x} = 
  \textbf{y}~\text{and}~
  \textbf{x}\in\Omega_{c}
 \end{cases},
 \label{clusterOpterator}
\end{align}
where $\overline{\textbf{M}}_{\textbf{x}}$, at any point in
$\Omega_{c}$, is some any linear operator ($3\times 3$ matrix) on the 
three-dimensional vector space at the point $\textbf{x}$. 
Recalling that $\mathbb{I}_{s}$ is defined as 
\begin{align}
 \mathbb{I}_{s}\left(\textbf{x},\textbf{y}\right) =
 \begin{cases}
  \overline{\textbf{0}} & \textbf{x} \neq \textbf{y}~\text{or}~
  \textbf{x}\not\in\Omega_{s} \\
  \overline{\textbf{1}} & \textbf{x} = 
  \textbf{y}~\text{and}~
  \textbf{x}\in\Omega_{s}
 \end{cases},
 \label{scattererProjection}
\end{align}
with $\overline{\textbf{1}}$ the identity operator ($3\times 3$ 
identity matrix), and that the composition of any two spatial 
projections is equivalent to a single spatial projection into the 
intersection of the two regions, any operator of the form given by 
Eq.~\eqref{clusterOpterator} commutes with $\mathbb{I}_{s}$, 
regardless of the actual geometry of the scatterer, allowing 
Eq.~\eqref{Tformal} to be rewritten as 
\begin{align}
 &\mathbb{T}_{s}^{\dagger}\mathbb{P}^{\dagger}_{_{\Omega_{c}}}
 \mathbb{I}_{s}
 =\mathbb{T}_{s}^{\dagger}\mathbb{P}^{\dagger}_{_{\Omega_{c}}}
 \mathbb{I}_{s}\left(\mathbb{V}^{-1}-\mathbb{G}^{\text{0}}\right)
 \mathbb{T}_{s}
 \nonumber \\
 &\Rightarrow\mathbb{T}_{s}^{\dagger}\mathbb{I}_{s}
 \mathbb{P}^{\dagger}_{_{\Omega_{c}}}
 =\mathbb{T}_{s}^{\dagger}\mathbb{P}^{\dagger}_{_{\Omega_{c}}}
 \mathbb{I}_{s}\left(\mathbb{V}^{-1}-\mathbb{G}^{\text{0}}\right)
 \mathbb{T}_{s}
 \nonumber \\
 &\Rightarrow\mathbb{T}_{s}^{\dagger}
 \mathbb{P}^{\dagger}_{_{\Omega_{c}}}
 =\mathbb{T}_{s}^{\dagger}\mathbb{P}^{\dagger}_{_{\Omega_{c}}}
 \left(\mathbb{V}^{-1}-\mathbb{G}^{\text{0}}\right)
 \mathbb{T}_{s}
 \nonumber \\
 &\Rightarrow \mathbb{P}_{_{\Omega_{c}}}\mathbb{T}_{s} =
 \mathbb{T}^{\dagger}_{s}\mathbb{U}\mathbb{P}_{_{\Omega_{c}}}
 \mathbb{T}_{s}.
 \label{mfFormal}
\end{align}
Working with increasingly refined collections of clusters, sets 
of subdomains $\left\{\Omega_{c}\right\}$ with $c$ ranging over
some indexing set such that $\cup_{c}\Omega_{c} = \Omega$, 
Eq.~\eqref{mfFormal} leads to increasingly refined mean-field 
approximations that mirrors the ``bottom up'' optimization of a
given objective with respect to structural degrees of 
freedom~\footnote{Cluster refinements, as shown in 
Ref.~\cite{molesky2020hierarchical}, can be given a mathematical 
ordering that is preserved in the limit values obtained in the 
associated optimization, whether or not the dual is used to relax 
the optimization.}. 
Letting $\left|\textbf{E}\right>$ denote some predefined electromagnetic 
flux, and making the simplifying choice that 
$\overline{\textbf{M}}_{\textbf{x}} = \overline{\textbf{1}}$ for all 
$\textbf{x}\in\Omega_{c}$ acting on Eq.~\eqref{mfFormal} with 
$\big<\textbf{E}\big| \ldots \big|\textbf{E}\big>$ gives 
\begin{align}
 &\int\limits_{\Omega_{c}}d\textbf{x}~
 \left[\textbf{E}^{*}\left(\textbf{x}\right)\cdot
 \textbf{T}\left(\textbf{x}
 \right) -
 \int\limits_{\Omega}d\textbf{y}~
 \textbf{T}^{*}\left(\textbf{y}\right)\cdot\overline{\textbf{U}}
 \left(\textbf{y},\textbf{x}\right)\cdot
 \textbf{T}\left(\textbf{x}\right)\right]
 \nonumber \\
 &=\big<\textbf{E}\big|\mathbb{P}_{_{\Omega_{c}}}
 \big|\textbf{T}\big> - \big<\textbf{T}\big|
 \mathbb{U}\mathbb{P}_{_{\Omega_{c}}}\big|\textbf{T}\big> = 0,
 \label{mfSingleVec}
\end{align}
where the generated polarization current $\big|\textbf{T}\big>$ is 
the image of $\big|\textbf{E}\big>$ under the action of 
$\mathbb{T}_{s}$ ($\mathbb{T}_{s}\big|\textbf{E}\big> \mapsto 
\big|\textbf{T}\big>$).
Returning to Eq.~\eqref{mfSingleVec}, the selection of any cluster 
$\Omega_{c}$ therefore defines an averaging kernel for the scattering 
theory: over $\Omega_{c}$ the otherwise free field must respect 
the true physical equations of electromagnetism, however, at any 
point $\textbf{x}\in\Omega_{c}$ unphysical fluctuations of the field 
(deviations of the integrand from zero) are permitted~\footnote{In 
particular, if $\Omega_{c} = \Omega$, so that there is only a single 
cluster filling the entire domain, then the symmetric 
($\sym{\mathbb{A}} = \left(\mathbb{A} + 
\mathbb{A}^{\dagger}\right) /2$) and antisymmetric 
($\sym{\mathbb{A}} = \left(\mathbb{A} + 
\mathbb{A}^{\dagger}\right) /2$) parts of the scattering constraints 
impose that both real and reactive power must be conserved, as was 
studied in Refs.~\cite{molesky2019bounds,gustafsson2019upper}.}.

As any physical polarization current field $\big|\textbf{J}^{g}\big>$ 
will automatically satisfy any relation of the form of 
Eq.~\eqref{mfSingleVec}, any optimization that only imposes such 
relations over a finite set of clusters $\left\{\mathbb{P}_{k}\right\}$ 
will automatically possess objective values at least as optimal as what 
can be achieved if Eq.~\eqref{Tformal} is imposed in full. 
Hence, solving the mean-field optimization 
\begin{align}
 &\text{min / max}_{~\left|\textbf{T}\right>}~\text{Obj}
 \left(\big|\textbf{T}\big>\right)
 \nonumber \\
 &\text{such that}~\left(\forall~\Omega_{k}\right)~
 \big<\textbf{E}\big|\mathbb{P}_{_{\Omega_{k}}}\big|\textbf{T}\big> 
 -\big<\textbf{T}\big|\mathbb{U}\mathbb{P}_{_{\Omega_{k}}}\big|
 \textbf{T}\big>,
 \label{mfOptProb}
\end{align}
similar to the mean-field theories used in statistical
physics~\cite{plefka1982convergence,jin2016cluster}, game
theory~\cite{nourian2013e,sen2016mean} and
machine-learning~\cite{kappen1998boltzmann,chen2018dynamical}, results
in a bound (limit) on physically realizable
performance~\footnote{Duality can also be used to solve a relaxation
  of the mean-field optimization that in many cases is in fact a true
  solution~\cite{boyd2004convex}.}.  Due to the implicit wavelength
scale contained in the Green's function ($\mathbb{G}^{\text{0}}$ in
$\mathbb{U}$), there is, for most objectives, no meaningful difference
between a mean-field fluctuating sufficiently rapidly and a physical
field, satisfying Eq.~\eqref{Tformal} exactly.  As a consequence,
smaller clusters, effectively, lead to higher order mean-field
theories that more closely bound what is actually possible.  This
tightening with decreasing cluster size causes \eqref{mfOptProb} to
act as an intriguing complement to standard structural design.  In
``bottom-up'' geometric optimization, the introduction of additional
degrees of freedom opens additional possibilities that improve
achievable optima.  In ``top-down'' mean-field optimization, the
introduction of additional clusters results in additional constraints
that reduce the space of possibilities available to the design field
$\big|\textbf{T}\big>$.  Our present high-level approach to solving
Eq.~\eqref{mfOptProb} is described in
Ref.~\cite{molesky2020hierarchical}.

\subsection{Multiple transformations (channels)}
\label{multtrans}

When handling a collection of sources that span a relatively small 
subspace, an efficient approach to computing $\mathbb{T}$ 
operator bounds through Eqs.~\ref{fluxInOut}~\&~\ref{mfFormal} 
is to work with individual inputs and outputs. 
Take $\left\{\left|\textbf{S}_{k}\right>\right\}$ to be a given 
collection of sources, and let $\left\{\left|\textbf{T}_{k}\right>
\right\}$ be the collection of polarization fields 
resulting from the action of $\mathbb{T}_{s}$, $\mathbb{T}_{s}\left|
\textbf{S}_{k}\right>\mapsto\left|\textbf{T}_{k}\right>$. 
Following Ref.~\cite{molesky2020hierarchical}, for any pair of indices 
$\left<k_{1},k_{2}\right>$, and each $\mathbb{P}_{_{\Omega_{c}}}$ 
cluster operator, $\left|\textbf{S}_{k_{1}}\right>$ and 
$\left\{\left|\textbf{T}_{k_{1}}\right>,\left|
\textbf{T}_{k_{2}}\right>\right\}$ must obey the relation 
\begin{equation}
 \left<\textbf{S}_{k_{1}}\right|\mathbb{P}_{_{\Omega_{c}}}
 \left|\textbf{T}_{k_{2}}\right> = 
 \left<\textbf{T}_{k_{1}}\right|\mathbb{U}\mathbb{P}_{_{\Omega_{c}}}
 \left|\textbf{T}_{k_{2}}\right>,
 \label{constraintsMatrix}
\end{equation}
where, again, $\mathbb{U} =\mathbb{V}^{-1\dagger} -
\mathbb{G}^{\text{0}\dagger}$, and $\left<\textbf{F}|\textbf{G}\right>$ 
denotes the standard complex-conjugate inner product over 
the complete domain ($\left<\textbf{F} | \textbf{G}\right> = 
\int_{\Omega}d\textbf{x}~\textbf{F}\left(\textbf{x}\right)^{*}
\cdot \textbf{G}\left(\textbf{x}\right)$ in the spatial basis). 
In Eq.\eqref{constraintsMatrix}, the extension to pairs of sources 
and polarization fields, compared to the single source constraints 
examined in Ref.~\cite{molesky2020hierarchical}, is necessary to 
account for the fact that a single scattering object (structured 
media) simultaneously generates each $\left|\textbf{T}_{k}\right>$ 
from each $\left|\textbf{S}_{k}\right>$. 
Over some set of clusters, using only the ``diagonal'' constraint 
between each source field ($\left|\textbf{S}_{k}\right>$) and 
polarization response ($\left|\textbf{T}_{k}\right>$) is equivalent to 
considering each source independently. 
But, by additionally including ``off-diagonal'' interactions between 
pairs, Eq.~\eqref{constraintsMatrix} introduces the requirement of a 
single consistently defined scattering object. 
Supposing a spatial basis in the limit of ``point'' (vanishingly 
small) clusters and complete field mixing, 
Eq.~\eqref{constraintsMatrix} becomes 
$\left(\forall~ x\in\Omega~\&~\left<k_{1},k_{2}\right>\right)$ 
\begin{align}
 &\textbf{S}_{k_{1}}^{*}\left(\textbf{x}\right)\cdot
 \overline{\textbf{M}}_{\textbf{x}}\cdot
 \textbf{T}_{k_{2}}\left(\textbf{x}\right) 
 \nonumber \\
 &=\left(\int\limits_{\Omega}d\textbf{y}~\textbf{T}^{*}_{k_{1}}
 \left(\textbf{y}\right)\cdot 
 \overline{\textbf{U}}\left(\textbf{y},\textbf{x}
 \right)\right)\cdot\overline{\textbf{M}}_{\textbf{x}}\cdot 
 \textbf{T}_{k_{2}}\left(\textbf{x}\right).
 \label{spatialConstraint}
\end{align}
Therefore, making use of the possible freedom in the definition of 
$\overline{\textbf{M}}_{\textbf{x}}$, whenever any 
$\textbf{T}_{k_{n}}\left(\textbf{x}\right)$ is nonzero any of the three 
local vector coordinates we must have 
\begin{equation}
 \textbf{S}_{k}^{*}
 \left(\textbf{x}\right) = \int\limits_{\Omega_{s}}
 d\textbf{y}~\textbf{T}^{*}_{k}\left(\textbf{y}\right) \cdot
 \overline{\textbf{U}}\left(\textbf{y},\textbf{x}\right),
 \label{pointInverse}
\end{equation}
for all $k$ and $\textbf{x}\in \Omega_{s}$, the collection of all 
spatial points in $\Omega$ where some 
$\textbf{T}_{k}\left(\textbf{x}\right) \neq \textbf{0}$. 
Taking the adjoint, the totality of the point constraints given by 
Eq.~\eqref{pointInverse} can be codified as 
\begin{equation}
 \left|\textbf{T}_{k}\right> = 
 \left(\mathbb{U}^{\dagger}_{_{\Omega_{s}}}\right)^{-1}
 \left|\textbf{S}_{k}\right>. 
\end{equation}
Comparing with Eq.~\eqref{Tformal}, $\Omega_{s}$ thus determines the 
unique geometry of the scattering object (structured medium) for all 
sources. 
\section{Applications}
\label{applications}

In this section, the ideas presented in Sec.~\ref{techDis} are applied
to several example applications, expounding the use of input--output
language to describe physical design as QCQP optimization problems,
and demonstrating the utility and tightness of the associated bounds.
Three main overarching features, which credibly apply beyond the
actual scope studied, are seen. First, distinct channels can display
widely varying (often unintuitive) characteristics. Second, both in
terms of computed bounds and the findings of inverse design,
attainable performance may vary greatly as the number of channels
increases---the design freedom offered by spatial structuring in a
wavelength scale device is by no means inexhaustible. Third, increasing 
the number of constraint clusters can improve the accuracy of bounds
calculations, but this improvement is problem dependent and varies, in
particular, with the material assumed. Throughout, as no feature size
can be imposed if one wishes to set bounds on any possible geometry,
quoted inverse design values represent the performance of
``grayscale'' structures, allowing each pixel to take on any
susceptibility value of the form $t\chi$ with $t\in\left[0,1\right]$.

\subsection{Field screening}

\emph{Objective}---Minimize the spatially integrated field intensity 
over a subwavelength ball for two dipole fields, supposing that the 
distribution of scattering material (of a predefined susceptibility) 
must be contained in a thin surrounding shell. 

The observation (shielded) region is taken to have a 
radius of $0.2~\lambda$, while the design domain is confined within a 
shell of inner radius $0.2~\lambda$ and outer radius $0.24~\lambda$. 
As sketched in the inset of Fig.~3(a), the ``separation vectors'' 
connecting the center of each dipole sources to the center of the 
observation domain are supposed to be aligned. 
The loss objective is defined as 
\begin{align}
 \text{Loss}\left(\big|\textbf{E}^{t}_{a}\big>
 ,\big|\textbf{E}^{t}_{b}\big>\right) = 
 \frac{\lVert \mathbb{I}_{\text{obs}}
 \big|\textbf{E}^{t}_{a}\big> \rVert_{2}^{2} + 
 \lVert \mathbb{I}_{\text{obs}}\big|\textbf{E}^{t}_{b}\big> 
 \rVert_{2}^{2}}{
 \lVert \mathbb{I}_{\text{obs}}\big|\textbf{E}^{i}_{a}\big> 
 \rVert_{2}^{2} + 
 \lVert \mathbb{I}_{\text{obs}}\big|\textbf{E}^{i}_{b}\big> 
 \rVert_{2}^{2}
 }
 \label{shieldObj}
\end{align}
with $\lVert \dots \rVert_{2}$ denoting the Euclidean two-norm and 
$\mathbb{I}_{\text{obs}}$ projection into the observation domain. 
Two clusters, evenly dividing the radius of the design region 
($0.2\rightarrow 0.22~\lambda$ and $0.22\rightarrow 0.24~\lambda$), 
are used in all bound computations. 

\begin{figure}[t!]
 \centering
 \includegraphics[width=1.0\columnwidth]{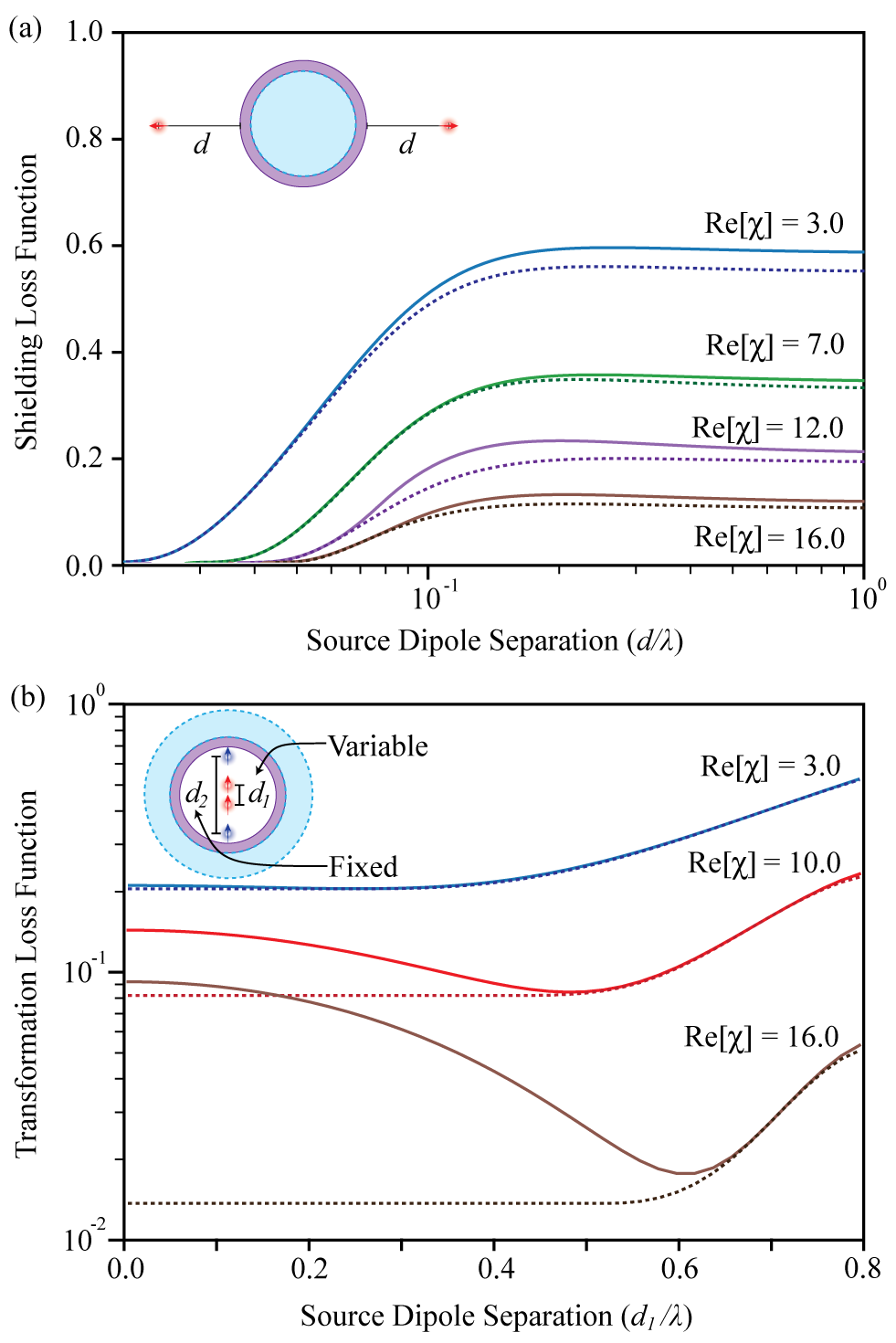}
 \caption{\textbf{Dipole screening and field transformations.} The
  figure depicts two representative three-dimensional applications
  of input--output bounds: limits on the ability of any structured
  geometry confined to a thin shell to (a) screen dipole fields, and
  (b) mask the separation of dipole source pairs within the
  encompassing environment. In both situations, dashed lines mark
  independent bounds, calculated without the inclusion of
  cross-constraints which enforce a unique optimal structure across
  all field transformations, while the solid lines result when
  source interactions effects are taken into account. A material
  loss value of $\im{\chi} = 0.01$ is assumed for all cases. }
\end{figure}

In the limit of large separation, the dipole fields within the design
and observation domains closely resemble counter-propagating
planewaves. Hence, the constant asymptote value seen as $d$
approaches $\lambda$ is, for all practical purposes, a bound on
shielding for two opposing planewaves. The decrease of the bounds for
small separation, speculatively, is caused by a combination of 
localization effects. As the dipoles are brought into the near-field of 
the design, the bulk of the field magnitude increasingly shifts towards
the edge of the observation region. Consequently, inducing small
shifts in the position of the field results in a comparatively larger
field expulsion. The presence of relatively larger evanescent fields
likely also simplifies the suppression of total emission via the
(inverse) Purcell effect.

The imposition of cross-constraints (solid lines), which in the limit
of point clusters enforce that all considered channels are realized
within one common geometry (see Sec.~\ref{multtrans}), is seen to 
produce a relatively small correction to the bounds attained
for independently optimized channels (dashed lines). This suggests
that although perfect shielding is not possible at sufficiently large
dipole separation, given the small shell and material constraints
investigated, many structures may potentially achieve similar
performance, i.e. the considered channels have yet to ``set'' the
scattering structure in any meaningful way.

\subsection{Dipole resolution and masking}

\emph{Objective}---Manipulate the field emanating from a pair of 
axially aligned dipoles, again limiting the volume available for device 
design to a thin encompassing spherical shell, so that exterior to the 
shell it appears that the dipoles are separated by either a larger or 
smaller distance. 
If the field is altered so that the separation distance appears 
larger than it actually is, it is easier to resolve that there are in 
fact two dipoles present, instead of a single dipole of some 
effective strength. 

More exactly, the input field is generated by a pair of $\textbf{z}$
polarized dipole sources, labeled 1 and 2, situated at $+d_i\unitv{z}$
and $-d_i\unitv{z}$ for $i=1,2$, and the target field is taken to
arise from equivalent dipolar configuration with a distinct
separation, $\pm d_t\unitv{z}$. The observation domain is set to be a
shell of half wavelength thickness enveloping the design domain, which
extends from an inner radius of $0.48~\lambda$ to an outer radius of
$0.5~\lambda$. 
Bounds, in all cases, are computed using ten spherical clusters, 
evenly spanning the design domain with respect to radius. 
Additional technical details concerning the example are given in 
Sec.~\ref{appB}.

\begin{figure*}[t!]
 \centering
 \includegraphics{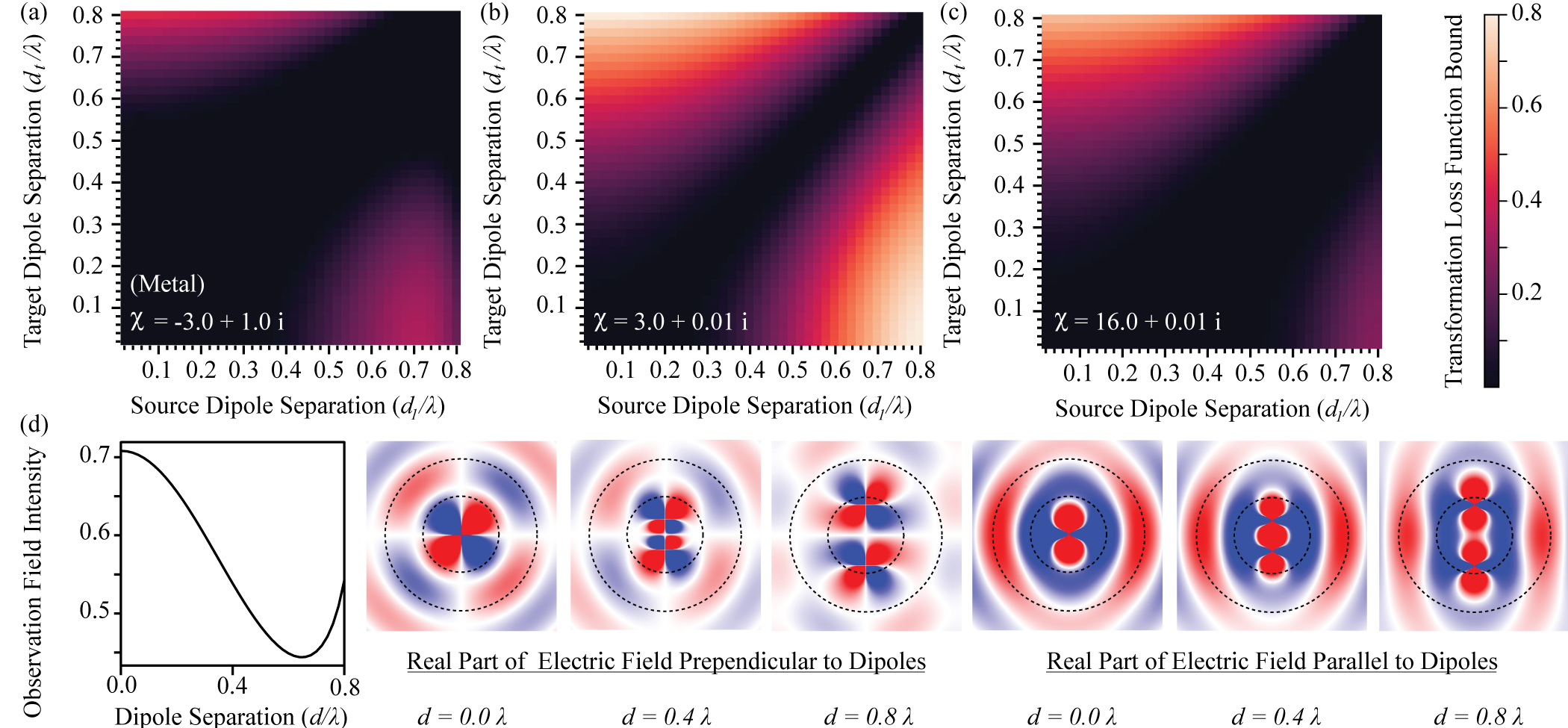}
 \caption{\textbf{Dipole resolution and masking.} Considering the
  same three-dimensional system depicted in the inset of Fig.~3(b),
  but with a single dipole-pair source, the three heat maps depict
  bounds on the ability of an arbitrary geometry made from a
  specified material, restricted to a thin shell, to modify the 
  perceived separation of a pair of axially aligned dipole sources, 
  as implied by the electric field exterior to the design. 
  Setup details are given in the main text.
  In moving from the trivial diagonal (no transformation) to the top
  left, the target field is set to be the field of an axial dipole
  pair further separated than that of the source. Opposingly, in
  moving to the bottom right, the target field is set to be the
  field of an axial dipole pair less separated than that of the
  source. 
  The observed asymmetry between these two directions provides 
  evidence of the ostensibly greater challenge of resolving
  closely separated fields, compared to masking separation. 
  Field profiles for representative dipole pairs, along with the average 
  field intensity within the observation domain (dashed lines), 
  are given in panel (d).}
\end{figure*}

The bounds depicted in the three heat maps of Fig.~4, exploring
variations in both input $d_{1} = d_{2}$ and output $d_{t}$ pair
separations, indicate that the challenge of either reducing or
enlarging the perceived separation of the source varies sizably with
supposed material properties and gap distances. 
For larger target or source separations (dipoles approaching the 
interior surface of the design domain) the increasingly dominant role of 
evanescent fields makes almost any transformation challenging, 
especially in weakly polarizable media. 
Focusing on Fig.~4(b), where the real part of the
permittivity is limited to small positive numbers (weak dielectrics),
the bounds prove that neither separation manipulation behavior is
achievable in any practical sense when the input and target
separations vary by more than $\approx 0.25~\lambda$. Conversely,
even under the imposition of considerably larger material loss
($\im{\chi}$), panel (a) suggests that much better performance is
achievable with metals ($\re{\chi} < -1$). 
Intuitively, the subwavelength characteristics of the system do not 
preclude resonant response in this case since plasmon excitations 
remain possible. 
The observed asymmetry between increasing and reducing perceived
separation seen in panels (b) and (c), which is largely 
absent in the metallic example of panel (a), is likely also tied to 
decaying waves: larger gaps produce larger evanescent fields in the 
observation domain compared to what is provided by the source, but 
smaller target separations do not.

Figure~3(b) explores the impact of distinct source separations,
varying $d_1$ while keeping $d_{2} = 0.7~\lambda$ fixed, for a common
target field corresponding to two co-located dipoles at the center of
the ball, $d_t = 0$. The findings highlight the need to enforce that
performance bounds for multiple channels refer to a unique scattering
geometry, achieved via the presence of cross-constraints in the
optimization (see Sec.~\ref{multtrans}). When the two channels are
analyzed separately (dashed lines), the bound values match those
displayed on the heat maps of Fig.~4, with the contribution of source
1 to the loss function approaching zero as $d_1 \to 0$, since in this
regime the source and target fields match. However, when
cross-constraints are included, the lack of a need for a design in
this regime for the first source is at odds with what is needed to
convert the field of the widely separated $d_{2}$ source, leading to
degraded performance.  As $d_{1}$ approaches $d_{2}$, the requirement
of a unique gemometry no longer introduces additional requirements,
and thus the ``simultaneous'' (cross-constrained) and ``independent''
results merge.

\subsection{Math kernels (integration and differentiation)}
\emph{Objective}---Within a bounding rectangle, design a
two-dimensional scattering profile, for a known collection of incident
waves, such that the relation between each incident field and total
field, along specified input and output observation planes, reproduces
the action of Volterra (integration) or differentiation operator.
Given that the Volterra and differentiation kernels are two of the
basic elements of differential equations, these examples are of
considerable interest in relation to recent proposals for optical
computing~\cite{estakhri2019inverse,li2019intelligent,
 rajabalipanah2020space}.

\begin{figure*}[t!]
 \centering
 \includegraphics{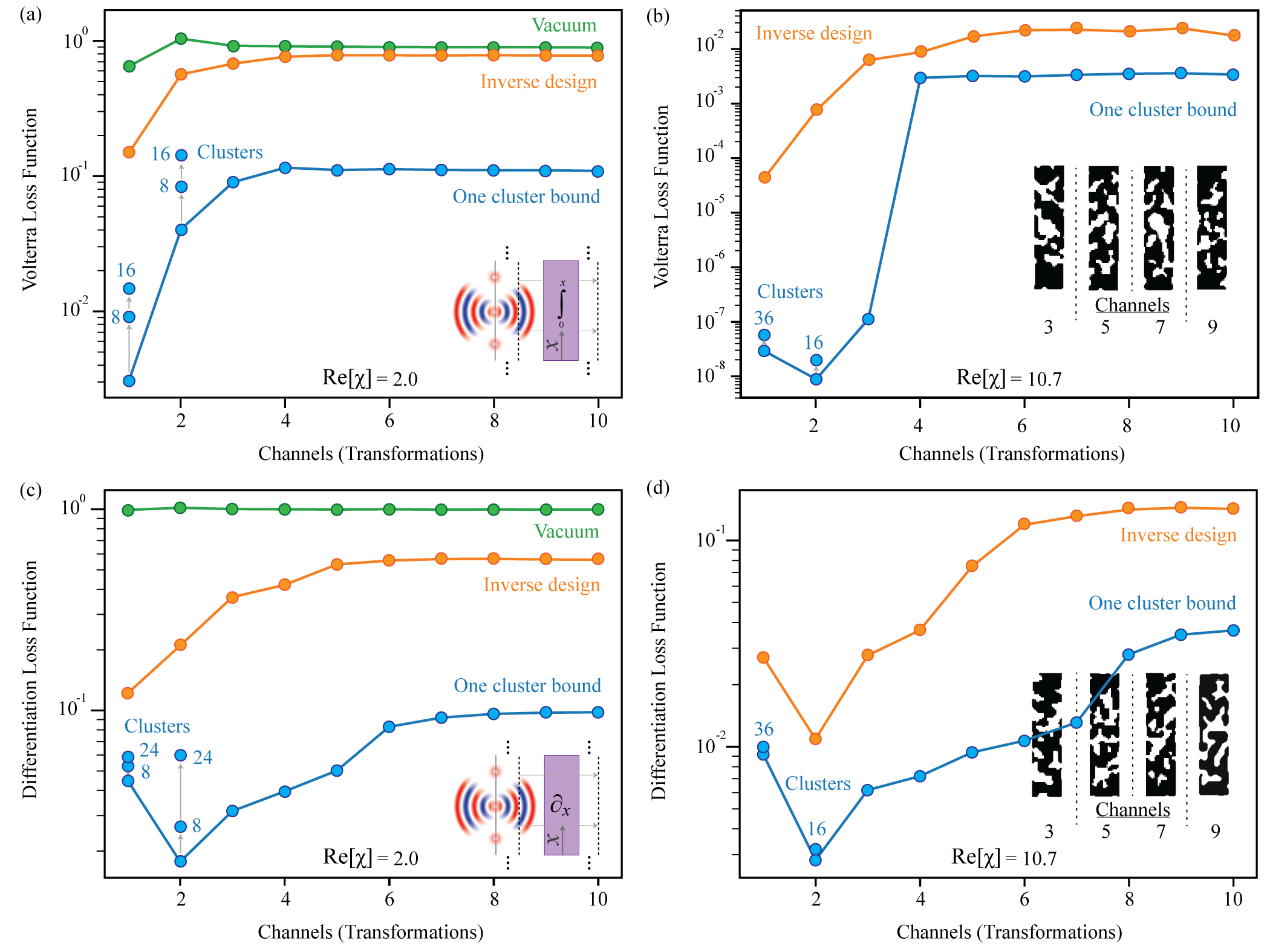}
 \caption{\textbf{Volterra and differentiation kernels.} Following the
   description in the main text, the figure depicts limits on the
   degree to which a structured medium of a given electric
   susceptibility $\chi$, confined to a rectangular design domain, may
   implement Volterra (top row) and differentiation (bottom row)
   kernels, as as function of the size of the subbasis (number of
   channels or transformations) over which the operator is defined.
   The cluster labels reference the number of cluster constraints
   imposed on the design domain, following the discussion of
   Sec.~\ref{techDis} and with further details given in the main
   text. Notably, the calculated bounds generally agree to within an
   order of magnitude of the performance of inverse designed
   structures when either the rank of the specified subbasis, or
   number of imposed clusters, is large.  In all four cases, a
   material loss value of $\im{\chi} = 0.01$ is
   assumed. Representative topology-optimized structures are included
   as insets in (b) and (d). The performance of these binarized
   systems is roughly a factor of two worse than the associated
   ``grayscale'' geometries compared to in the figure.}
\end{figure*}

The design region is chosen to have a transverse length (thickness) of
$0.5~\lambda$ and parallel length of $2~\lambda$, insets Fig.~5(a) 
and Fig.~5(c). 
The input observation plane is set at distance of $0.25~\lambda$ from 
the input edge of the design region, and the output observation plane 
is located at a distance of $0.25~\lambda$ from the output edge of the 
design region. 
Incident fields are generated by dipole (pixel) sources
placed on a source plane, running parallel to the long edge of the
design region, situated at a distance $0.2~\lambda$ from the input
observation plane. The source plane has length of $1.5~\lambda$ and
is centered in relation to the design domain, creating a
$0.25~\lambda$ offset from the top and bottom edges of the design
region and input and output planes. For the Volterra operator, the
target fields along the output plane, letting $y$ denote the direction
running parallel to the long edge of the design region, are calculated
as $\textbf{E}^{\diamond}\left(x_{\diamond}, y\right) =
\int^y_{0}\text{d}y'~\textbf{E}^{i}(x_{i},y')$, with $x_{\diamond}$
and $x_{i}$ denoting the common coordinate of the output and input
plane, respectively. For the differential operator, keeping the above
notation, the target fields are calculated as $\textbf{E}^{\diamond}
\left(x_{\diamond}, y\right) = \partial \textbf{E}^{i}\left(x_{i},
y\right)/\partial y$, implemented as a finite-difference approximation
on a computational grid. The optimization objective is defined by
Eq.~\eqref{fluxAgree} and the bounds formulated following the
description given in Sec.~\ref{techDis}. Further implementation
details appear in Sec.~\ref{appC}. The loss function values appearing
in Fig.~5 are determined by
\begin{equation}
 \text{Loss}\left(\big|\overline{\textbf{E}^{t}}\big>\right) = 
 \frac{\lVert \mathbb{I}_{\text{out}}
 \left(\big|\overline{\textbf{E}^{t}}\big> - 
 \big|\overline{\textbf{E}^{\diamond}}\big>\right)
 \rVert^{2}_{2}}{\lVert
 \mathbb{I}_{\text{out}}\big|\overline{\textbf{E}^{i}}\big>
 \rVert^{2}_{2} + 
 \lVert\mathbb{I}_{\text{out}}
 \big|\overline{\textbf{E}^{\diamond}}\big>
 \rVert^{2}_{2}}.
 \label{mathLossFunc}
\end{equation}
In Eq.~\eqref{mathLossFunc}, $\mathbb{I}_{\text{out}}$ denotes 
projection onto the output observation plane, and an overline on a 
vector indicates vertical concatenation over all like named fields 
indexed by the incident waves, i.e. $\big|
\overline{\textbf{E}^{t}}\big> = \left[\big|\textbf{E}^{t}_{1}\big>~,
\dots,~\big|\textbf{E}^{t}_{_{N}}\big>\right]$ for $N$ sources. 
The cluster numbers appearing in Fig.~5 refer to specific, even and 
consistent, divisions of the design domain rectangle along its long 
and short edges. 
Writing division of the short edge first, $8 = 2\times 4$, 
$16 = 2\times 8$, $24 = 3 \times 8$ and $36 = 3 \times 12$. 

The most striking aspect of the findings displayed in Fig.~5,
especially when viewed in conjunction with Fig.~6, is the widely
varying characteristics observed between the four cases. 
For the Volterra and differentiation kernels, at least within the 
subbasis used here, the number of channels considered has a remarkable
influence on the observed limit values.
Given that analogous trends are seen in performance of the 
associated structural optimizations, it can be safely concluded that 
this behavior is fundamental, hinting that there is a sort of 
Fourier (resolution) limit at play (e.g., size-dependent constraints 
on space-bandwidth products are known to limit the degree to which a 
given optical feature may be resolved with a finite 
device~\cite{mendlovic1997space,neifeld1998information}). 
Moreover, presumably because differentiation demands creation of more 
rapid profile variations whereas Volterra operations demand only 
smoothing, loss function values (for both bounds and inverse design) 
differ by multiple orders of magnitude for small numbers of channels 
($\approx 3$) before reaching ``asymptotes'' that differ roughly by a 
factor of $10$, for $\chi = 10.7 +i~0.01$. 
Conversely, in the focusing example examined in Fig.~6, the bounds 
exhibit only a weak dependence on the number of channels in either 
representative material. 
While a precise account of the factors leading to these differences 
is likely challenging, it should be noted that in all cases the
objective is bounded (from above or below) by performance that could
be achieved in vacuum or a full slab, and so the appearance of
plateaus when performance is poor should come as no surprise. 
A related, albeit more intuitive, dichotomy is seen with respect to the
number of clusters imposed and the magnitude of electric
susceptibility ($\re{\chi}$). For $\re{\chi} = 2$, subdividing the
domain generally leads to substantial tightening, while for $\re{\chi}
= 10$ many more clusters are needed to produce meaningful alterations
owing to the decrease in the effective wavelength in the medium.

\subsection{Near-field lens (metaoptics)}
\emph{Objective}---Working within a given bounding rectangular design
domain, maximize the average field intensity within a specified focal
region for a predefined set of incident plane waves spanning a cone of
incidence.

The size of the rectangular design region is characterized by the
lengths $L_{1}^{d}=0.5~\lambda$ and $L_{2}^{d}=2~\lambda$, inset
Fig.~6. The focal region is a square centered along $L_{2}^{d}$, with
side length $L_{1}^{f}=L_{2}^{f}=0.25~\lambda$. The middle of the
focal region is set to reside at a distance of $0.5~\lambda$ from the
outgoing edge of design region. The design objective is
\begin{align}
 \text{Obj}\left(\Big\{\big|\textbf{E}^{t}_{k}\big>\Big\}\right)=
 \text{max} \sum_{i=1}^N\big<\textbf{E}^{t}_{i}\big|
 \mathbb{I}_{\text{focal}}\big|\textbf{E}^{t}_{i}\big>
\end{align}
where the summation index runs over the $N$ angles of incidence,
evenly distributed over a cone of $(N-1)\times 15^\circ$ degrees
centered on the midpoint of $L^{2}_{d}$. As before, the cluster
labels given in Fig.~6 reference even divisions of the rectangular
design domain along its short and long edges: $16 = 2 \times 8$, $32 =
4\times 8$, $36 = 3\times 12$, $100 = 5 \times 20$, and $256 = 8\times
32$.

\begin{figure}[t!]
 \centering
 \includegraphics[width=1.0\columnwidth]{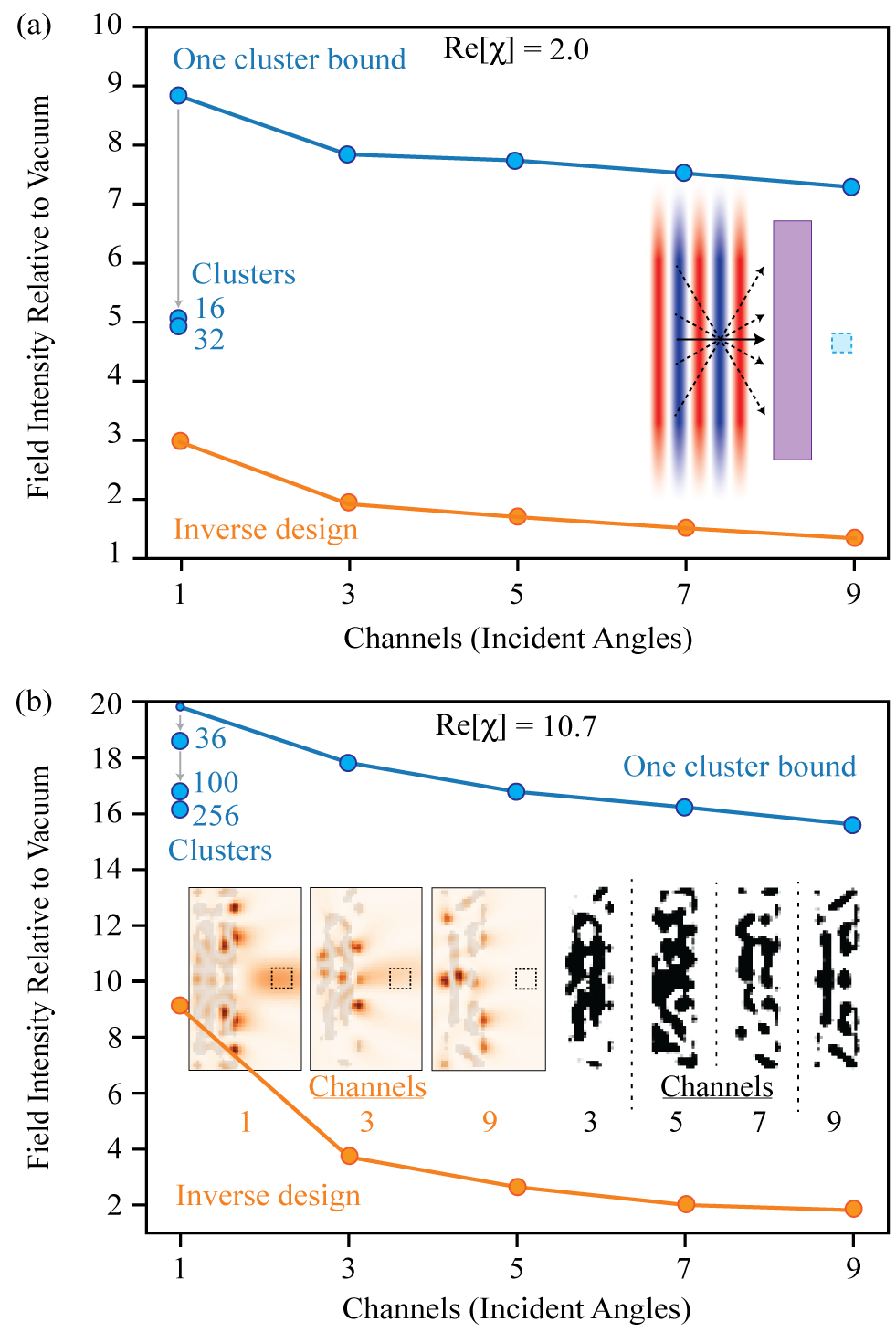}
 \caption{\textbf{Focusing metalens.} Highlighting the applicability
   of the proposed method to establish limits on metaoptical elements,
   the figure provides bounds on the ability of any structure confined
   within a rectangular domain of subwavelength thickness to enhance
   the average field intensity over a predefined focal region in the
   near field of the structure, for a given collection of incident
   planewaves. Also shown are comparative performances of inverse
   designed structures. A material loss value of $\im{\chi} = 0.01$ is
   supposed in all cases. Binarized structures for $3,~5,~7$ and $9$
   channels, producing relative field-enhancement factors of
   $5.27,~2.66,~2.56$ and $~1.64$ respectively, are included as
   insets. Quoted values for the inverse design curve correspond to
   ``grayscale'' structures.}
\end{figure}

Even in comparison to previous examples, the extent to which inverse
design is able to approach the limit values produced by the
formulation of Sec.~\ref{techDis} is notable. For a single source (a
normally incident planewave), agreement within a factor of $3$ is
achieved when a single cluster constraint is supposed, and as
increasingly localized domain constraints are imposed the performance
gap is seen to nearly close.  Hence, somewhat surprisingly, the
relatively poor field localization values achieved by the topology
optimized (inverse design) geometries are in fact not far from ideal.
We suspect that three underlying factors are largely responsible for
these findings.  First, the focusing problem does not impose any
specific target profile for the output fields, inset Fig.~6(b),
leading to a greater variety of geometric possibilities with similar
focusing performance.  Second, given the assumed design domain, the
specified location and volume of the focal region makes the problem
more difficult than what might be naively
expected~\cite{christiansen2021inverse}; quite different results may
be encountered if the ``spot size'', or separation between the focus
and design domain, are reduced.  Third, the extended nature of
planewave sources naturally suggests the uniform cluster constraints
that have been employed.  Contrastingly, for the Volterra and
difference kernels, it is plausible that non-uniform cluster
distributions, or simply smaller clusters, could lead to tighter
bounds.  A systematic study of such questions, once a more
resource-effective approach for solving Eq.~\eqref{optProb} is
found~\cite{angeris2021heuristic}, remains as an important direction
for future work.


\section{Summary discussion}

In summary, we have shown that recently developed methods for
calculating limits on sesquilinear electromagnetic
objectives~\cite{angeris2019computational,kuang2020computational,
  molesky2020hierarchical,jelinek2020sub,angeris2021heuristic}, can be
applied to a wide variety of design problems, including spatial
multiplexing~\cite{richardson2013space,li2017recent, yang2020density},
metaoptics~\cite{kruk2017functional,
  lewi2019thermally,staude2019all,lin2020end}, linear
computation~\cite{estakhri2019inverse,li2019intelligent,
  rajabalipanah2020space} and light
extraction~\cite{schneider2018numerical,shen2020broadband,
  wambold2021adjoint,chakravarthi2020inverse}. Employing the language
of communication theory, this program stands as a significant
extension of prior work on channel-based electromagnetic
limits~\cite{miller2007fundamental,miller2012all,
  miller2013self,miller2013complicated,miller2017universal,
  miller2015shape,pai2019matrix,miller2019waves}. Namely, although
highly insightful, the characteristics of the background Green's
function connecting the volumes containing particular sender and
receiver registers are generally insufficient for accurately assessing
whether the extent to which some desired communication can occur,
particularly in wavelength-scale devices. Rather, as may be confirmed
by a survey of the results presented in Sec.~\ref{applications}, the
degree to which communication between a predefined collection of
registers can occur may depend strongly on a range of other
environmental factors, such as the physical size and response
parameters available for designing the channels, and the spatial
profile of the register fields.

The strong correspondence observed between bounds computed with this
approach and the findings of inverse design exemplified in
Sec.~\ref{applications} continues many of the trends seen in the
earlier works cited above. Regardless of the particular objective and
distribution of cluster constraints considered, we have yet to
encounter a situation in which strong duality does not
hold~\footnote{It should be mentioned that we have encountered
  situations in which very high numeric precision was required to
  verify the presence of strong duality.}. As such, it would seem that
the outstanding difficulties of constructing a general toolbox for
realizing (tight) limits that incorporate the full wave physics and
limitations of Maxwell's equations, and in turn offer guidance for
practical designs, are computational~\cite{molesky2020t,
  molesky2020hierarchical,angeris2021heuristic}. All present evidence
indicates that optimization problems following the form of
Eq.~\eqref{mfOptProb} can be solved exactly via the duality relaxation
described in Sec.~\ref{techDis}. If this is indeed the case, then the
matter of central importance is not whether there is some
configuration of clusters that will capture all key effects that
physically limit communication, but whether some reasonably good
approximation can actually be solved in an acceptable amount of time.
With this in mind, we believe that the results presented above provide
substantial reason for optimism.

Finally, the specificity of the program presented in the main text
raises a prescient question pertaining to realizing abstract
transformations.  Expressly, there are many instances in which the
primary concern is how well the transformation can be implemented
between a given number of \emph{undetermined} inputs and outputs
(channels). The distinction amounts, in essence, to the same shift in
perspective advocated for in recent ``end-to-end'' inverse design
formulations~\cite{valmadre2017end,lin2020end} and waveform
optimization investigations~\cite{poyneer2005optimal,sitzmann2018end}: 
the exact characteristics of the domain and codomain sets for some 
particular discrimination or transformation are often mutable, and 
only important in so far as they implicitly alter achievable 
performance. The duality program outlined
above can be extended to treat this problem in a variety of ways, and
we intended to present this analysis in detail in an upcoming work.

\begin{acknowledgments}
This work was supported by the National Science Foundation under
the Emerging Frontiers in Research and Innovation (EFRI) program, 
EFMA-1640986, the Cornell Center for Materials 
Research (MRSEC) through award DMR-1719875, and the Defense Advanced 
Research Projects Agency (DARPA) under agreements HR00112090011, 
HR00111820046 and HR0011047197. 
The views, opinions and findings expressed herein are those of the 
authors and should not be interpreted as representing official 
views or policies of any institution. 
\end{acknowledgments}

\section{Appendix}

\subsection{Off-origin dipole spherical wave expansions}
\label{appB}

Following results of Ref.~\cite{wittmann1988spherical}, as explained 
in Refs.~\cite{xu1996calculation,kruger2012trace}, off-origin, on-axis, 
outgoing spherical waves can be expanded in terms of on-origin 
outgoing and regular spherical waves as
\begin{equation}
 \textbf{W}^{\text{out}}_{plm}(\textbf{r}) =
 \begin{cases}
  \sum\limits_{~p',l'} \mathcal{U}^{\text{in}\pm}_{p'p,l'lm}(d)~
  \textbf{W}^{\text{reg}}_{p'l'm}(\textbf{r}\pm d\unitv{z}) & r<d \\
  \sum\limits_{~p',l'} \mathcal{U}^{\text{out}\pm}_{p'p,l'lm}(d)
  ~\textbf{W}^{\text{out}}_{p'l'm}
  (\textbf{r}\pm d\unitv{z}) & r>d 
 \end{cases},
\end{equation}
where
\begin{widetext}
 \begin{align}
 &\mathcal{U}^{\text{in}\pm}_{p'p,l'lm}\left(d\right) 
 = \sum_\nu \bigg[ 
 \frac{l(l+1)+l'(l'+1)-\nu(\nu+1)}{2}
 \delta_{pp'} \mp i~m d k_{o}(1-\delta_{pp'})\bigg] 
 A^{h\pm}_{l'l\nu m}(d),
 \label{sphereFuncsFirst}
 \\
 &\mathcal{U}^{\text{out}\pm}_{p'p,l'lm}\left(d\right) 
 = \sum_\nu \bigg[\frac{l(l+1)+l'(l'+1)-\nu(\nu+1)}{2}
 \delta_{pp'} \mp i~m d k_{o}(1-\delta_{pp'})\bigg] 
 A^{j\pm}_{l'l\nu m}(d),
 \\
 &A^{h\pm}_{l'l\nu m}(d) = (-1)^m i^{l-l'\pm\nu} (2\nu+1)
 \sqrt{\frac{(2l+1)(2l'+1)}{l(l+1)l'(l'+1)}} 
 \begin{pmatrix}
  l & l' & \nu \\ 
  0 & 0 & 0
 \end{pmatrix}
 \begin{pmatrix}
  l & l' & \nu \\ 
  m & -m & 0
 \end{pmatrix}
 h_\nu\left(dk_{o}\right),
 \\
 &A^{j\pm}_{l'l\nu m}(d) = 
 (-1)^m i^{l-l'\pm\nu} (2\nu+1)
 \sqrt{\frac{(2l+1)(2l'+1)}{l(l+1)l'(l'+1)}} 
 \begin{pmatrix}
  l & l' & \nu \\ 
  0 & 0 & 0
 \end{pmatrix} 
 \begin{pmatrix}
  l & l' & \nu \\ 
  m & -m & 0
 \end{pmatrix}
 j_\nu\left(dk_{o}\right).
 \label{sphereFuncsLast}
\end{align}
\end{widetext}
In Eqs.~\eqref{sphereFuncsFirst}--\eqref{sphereFuncsLast}, 
the large round brackets are Wigner-3j symbols, $j_{\nu}$ and 
$h_{\nu}$ denote the spherical Bessel and Hankel functions of the 
first kind respectively, and the plus and minus signs indicate the sign 
of the necessary coordinate transformation. 
Noting that the field of a $\hat{\textbf{z}}$-polarized 
dipole radiation consists entirely of the $\textbf{N}_{1,0}$ 
outgoing wave,
\begin{equation}
 \textbf{E}^z(r,\theta,\phi) = \frac{ik_{o}}{\sqrt{6\pi}} 
 \textbf{N}_{1,0}\left(r,\theta,\phi\right),
 \label{zdipoleField}
\end{equation}
the field of an off-axis dipole can thus be computed from 
Eqs.~\eqref{sphereFuncsFirst}--\eqref{sphereFuncsLast} by setting 
$p=N,~l=1,$ and $m=0$ implying, through the Wigner selection rules that 
the on-origin regular $M$ waves, $\text{Rg}\textbf{M}$, waves 
need not be included. 
Using this knowledge, the expansion coefficients for an off-origin, 
on-axis, dipole can be reduce to 
\begin{align}
 \textbf{E}^{z}_{-d\hat{\textbf{z}}}(\textbf{r}) &= 
 \frac{ik_{o}}{\sqrt{6\pi}}\textbf{N}_{1,0}(\textbf{r}+d\unitv{z}) 
 \nonumber \\
 &=
  \begin{cases}
    \frac{ik_{o}}{\sqrt{6\pi}}\sum_{l=1}^{\infty} 
    \mathcal{U}^{\text{in}-}_{NN,l,1,0} \text{Rg}
    \textbf{N}_{l,0}(\textbf{r}) &\left(r<d\right),
    \\
    \frac{ik_{o}}{\sqrt{6\pi}}\sum_{l=1}^{\infty} 
    \mathcal{U}^{\text{out}-}_{NN,l,1,0}
    \textbf{N}_{l,0}(\textbf{r}) &\left(r>d\right),
  \end{cases}  
  \\
  \textbf{E}^{z}_{+d\unitv{z}}(\textbf{r}) &= 
  \frac{ik_{o}}{\sqrt{6\pi}}\textbf{N}_{1,0}(\textbf{r}-d\unitv{z}) 
  \nonumber \\
  &= 
  \begin{cases}
    \frac{ik_{o}}{\sqrt{6\pi}}
    \sum_{l=1}^{\infty} \mathcal{U}^{\text{in}+}_{NN,l,1,0} 
    \text{Rg}\textbf{N}_{l,0}(\textbf{r})& \left(r < d\right),
    \\
    \frac{ik_{o}}{\sqrt{6\pi}}
    \sum_{l=1}^{\infty} \mathcal{U}^{\text{out}+}_{NN,l,1,0} 
    \textbf{N}_{l,0}(\textbf{r})& \left(r > d\right),
 \end{cases}
\end{align}
where
\begin{align}
 &\mathcal{U}^{\text{in}-}_{NN,l,1,0} = 
 \sum_{\nu=l-1}^{l+1} L^{\nu}_{l}
 \begin{pmatrix}
  1 & l & \nu \\ 
  0 & 0 & 0
 \end{pmatrix}^2 
 h_\nu\left(dk_{o}\right),
 \\
 &\mathcal{U}^{\text{in}+}_{NN,l,1,0} = 
 \sum_{\nu=l-1}^{l+1}
 (-1)^\nu 
 L^{\nu}_{l}
 \begin{pmatrix}
  1 & l & \nu \\ 
  0 & 0 & 0 
 \end{pmatrix}^2
 h_\nu\left(dk_{o}\right).
\end{align}
$\mathcal{U}^{\text{out}-}_{NN,l,1,0}$ and 
$\mathcal{U}^{\text{out}+}_{NN,l,1,0}$ are equivalently defined but 
with the Hankel functions of the first kind, $h_\nu\left(dk_{o}\right)$, 
replaced by Bessel functions of the first kind. 
In these expressions, summation limits are set by explicit evaluation of the 
Wigner-3j selection rules and 
\begin{equation}
 L^{\nu}_{l}= 
 (2\nu+1)\frac{2+l(l+1)-\nu(\nu+1)}{2}
 i^{1-\nu-l}
 \sqrt{\frac{6l+3}{2l(l+1)}}.
\end{equation}

\subsection{Computation for two-dimensional examples}
\label{appC}

The Green's function representations used for computing bounds in all
two-dimensional examples were obtained by the open-source
\textit{ceviche} FDFD package~\cite{hughes2019Forward}.  The basis
used for representing fields in all such cases are the individual
discretization pixels of the FDFD computation.  Correspondingly, the
matrix representation of the Green's function connects every pixel
with a numeric approximation of the field generated by a dipole source
at the location of the pixel.
  
For all studies with a free space background, the translational 
invariance of the Maxwell is exploited to construct the matrix 
representation of the Green's function from a single dipole field 
solve. 
Concretely, consider a domain of shape $(N_x,N_y)$ in pixels. 
The required representation of $\mathbb{G}^{0}$ is then a square matrix of 
dimension $N_x N_y$, where the $j$-th column is the field 
created by an electric dipole at position $j$. 
This matrix can be obtained in a single solve by placing a dipole at 
the center of a larger $(2N_x -1, 2N_y -1)$ domain and observing the 
calculated field in sliding a window of size $(N_x, N_y)$. 

To account for the cross-source constraints in the presence of
multiple sources, compared to the approaches we have used in prior
works~\cite{molesky2020hierarchical}, the fields and matrices for each
of the individual sources are grouped into ``super'' vectors and
matrices:
\begin{align}
 &\big|\textbf{T}\big> = 
 \begin{bmatrix}
  \big|\textbf{T}_{1}\big> \\
  \vdots \\
  \big|\textbf{T}_{_{N}}\big>
 \end{bmatrix}
 &\mathbb{Z}^{_{TT}} = 
 \begin{bmatrix}
  \mathbb{Z}^{_{T_{1}T_{1}}} & \dots & \mathbb{Z}^{_{T_{1}T_{N}}}
  \\
  \vdots & \ddots & \vdots \\ 
  \mathbb{Z}^{T_N T_1} & \dots & \mathbb{Z}^{T_N T_N}
 \end{bmatrix}
\end{align}
As such, the dimension of the vectors and matrices involved in 
the calculation of a $N$-source bound scales linearly with $N$. 
To facilitate computation of bounds for large device footprints 
and values of $N$, an Arnoldi basis for $\mathbb{U}$ is computed to 
more efficiently represent the $\left|\textbf{T}\right>$ image field. 
For any given problem, the only vectors that interact with the 
primal degrees of freedom, $\left\{\big|\textbf{T}_{n}\big>\right\}$, 
are the sources, $\left\{\big|\textbf{E}^{i}_{n}\big>\right\}$, and 
the linear coefficients of $\left\{\big|\textbf{T}_{n}\big>\right\}$ 
in the primal objective 
$\left\{\big|\mathcal{O}^{\text{lin}}_{n}\big>\right\}$. 
Hence, these vectors
\begin{equation}
  B_{0} = 
  \begin{bmatrix}
   \big|\textbf{E}^{i}_{1}\big> & \dots & 
   \big|\textbf{E}^{i}_{_{N}}\big> & 
   \big|\mathcal{O}^{\text{lin}}_{1}\big>
   & \dots & 
   \big|\mathcal{O}^{\text{lin}}_{1}\big>
  \end{bmatrix},
\end{equation}
posses favorable convergence characteristic for 
computing the dual via Arnoldi iterations. 
For communications type problems 
$\big|\mathcal{O}^{\text{lin}}_{n}\big> = 
\mathbb{G}_{od}^{\dagger}\left(\big|\textbf{E}^{\diamond}_{n}\big> 
- \big|\textbf{E}^{i}_{n}\big>\right)$, 
while for the metalens problem 
$\big|\mathcal{O}^{\text{lin}}_{n}\big> 
= \mathbb{G}_{od}^{\dagger} \big|\textbf{E}^{i}_{n}\big>$. 
To begin the Arnoldi procedure, $B_{0}$ is orthonormalized to give 
$\bar{B}_{0}$. 
The $i$-th iteration is then computed by generating $B_{i} = 
\mathbb{U} \bar{B}_{i-1}$, and then orthonormalizing $B_i$ both 
internally and with respect to $\bar{B}_0, \dots, \bar{B}_{i-1}$, 
giving $\bar{B}_i$. 
The partial basis for the generated block Krylov subspace 
at the end of the $i$-th iteration is the thus the aggregate of all 
the column vectors of $\bar{B}_0, \dots, \bar{B}_i$. 
Denoting $\mathbb{U}$ restricted to the $i$-th Krylov subspace as 
$\mathbb{U}_i$, the representation is effectively complete when 
the column norms of $\mathbb{U}_{i}^{-1} \bar{B}_0$ converge 
to within a certain tolerance, Ref.~\cite{molesky2020t}. 

Inverse design results for all two dimensional examples are also
computed using \textit{ceviche} using a standard topology
optimization~\cite{christiansen2021inverse}.  For sake of comparison
with the optimization problem underlying the bound formulation,
grayscale structures are treated as viable.  That is, in any
particular design, the permittivity value of a given pixel can take
any value in the convex set bounded by $0$ and the quoted material
value.  


\end{document}